\documentstyle[12pt]{article}
\title{ Reduction of relativistic three-body kinematics}
\author{Ph. Droz-Vincent\\[2mm]LUTH\\
Observatoire de Meudon\\
5 place Jules Janssen\\
92195 Meudon, France}
   \date{ }
\newcommand  {\eeq}{\end{equation}}    
\newcommand  {\lam}{\lambda}    
 \newcommand {\dron}{\partial}
\newcommand  {\beq}{\begin{equation} }
\newcommand  \half {  {1 \over 2} }
\newcommand  {\ytil}{\widetilde y}
\newcommand{\yhat}{\widehat y}  
\newcommand {\ztil}{\widetilde z}
\newcommand{\zhat}{\widehat z}
\newcommand  {\noi}{\noindent}
\newcommand  {\disp}{\displaystyle}

\newcommand {\sig}{   \sigma   }
\newcommand {\alp}{\alpha}
\newcommand{\Gam}{\Gamma}
\newcommand{\soulD}{\underline D}
\newcommand{\soulV}{\underline V} 
\newcommand{\soulsig}{\underline \sigma}
\newcommand{\souleps}{\underline \epsilon}
\newcommand{\soulv}{\underline v}
\newcommand{\soulu}{\underline u}    
\begin{document}
\maketitle
\abstract{The Klein-Gordon system describing three scalar particles without 
interaction  is  cast into a new form, by transformation of the momenta.
 Two  redundant  degrees of  freedom are eliminated;
we are  left with a covariant   equation for a reduced wave function
 with three-dimensional arguments.
This new formulation of the mass-shell constraints is equivalent to
 the original KG system in a sector characterized by positivity of the energies 
and, if the mass differences are not too large, by a moderately  relativistic
  regime.\\ Introducing  mutual  interactions provides  a model which is
(at least for three equal masses)  tractable and admits a reasonable 
nonrelativistic limit.

\medskip


\bigskip  

\section{Introduction}

\subsection{Motivations}
\noi  Relativistic {\em particle\/} dynamics is concerned by situations where 
the  particles we consider are not significantly  created or anihilated,
 whereas other relativistic effects must be taken into account.
In principle the description of  such particles should result  from a 
specialization of quantum field theory (QFT) to its $n$-body sector.

\noi This line  leads to the famous integral equation of Bethe 
and Salpeter (BS) in the two-body case. Three-body generalizations have soon 
been considered in the litterature  \cite{JGTaylor}. More recently see 
\cite{Tjo}\cite{Bij}.
For $n>2$ however, the complexity of the BS approach seems to be  almost 
prohibitive as far as  practical applications are concerned.

\noi     An alternative approach,
 based upon first principles \cite{predic1} \cite{predic2}, 
uses   $n$ mass-shell constraints in the form of 
coupled wave equations where interaction terms can be either phenomenological 
or derived from  QFT  \cite{predic3}.  
          This method  shares with BS equation
 the property of   manifest relativistic invariance, realized at the price of
 dealing with redundant degrees of freedom, 
since the arguments of the wave function are four-vectors.
\noi In the two-body case, there is a clue for eliminating  the redundant  
 degree of freedom:   the sum of wave equations  rules 
the dynamics, whereas their difference allows to determine how the wave 
function depends on the  "relative time". This dependence turns out to be 
trivial and one is left with a three-dimensional problem.

\noi   In the three-body case we  have to cope with two
 "relative times".
 These superfluous degrees of freedom are  present as well in  the 
three-body versions of the BS equation. 
Their elimination (or factorization) is desirable for physical interpretation;
it would 
produce (after diagonalization of the total linear momentum) a reduced wave 
equation  which is covariant but similar to a  Schroedinger equation with
three-dimensional arguments.
Unfortunately,  the simple procedure  utilized in the two-body case does
 not work for  $n>2$.

\medskip
\noi     An  important issue of $n$-body dynamics is cluster separability; 
but a  less restrictive and  more essential requirement
is {\em global\/} separability: one must at least recover free-particle motion 
when {\em all} interactions are put equal to zero. Models violating 
 global  separability have been considered in the past \cite{llos}\cite{DV},
 mainly for their computational simplicity, but we belive that any reasonable
 formulation of  $n$-body dynamics must include free motion as a limit
 when all the terms carrying interactions are "switched off".

\noi
In sofar as fermions are concerned,  these matters have been discussed 
earlier in the literature \cite{bijNCim}  \cite{Saz1}.

\medskip
\noi   For scalar particles with masses $m_a$,
 free motion can be described by $n$ Klein-Gordon (KG) equations, say 
    $   (p_a^2 - m_a ^2) \   \Phi =0$    where $\Phi$ depends 
on the momenta $p_1, ....p_n$.
We can give a sharp timelike value $k^\alp $ to the total linear momentum
and use the differences of these equations.
In the two-body case, it follows that the relative time (or alternatively 
the relative energy 
$\disp  {c  \over 2}  (p_1 - p_2) \cdot  k /  \sqrt {k^2}$,
 which is conjugate to 
it)  arises only in a trivial factor of the wave function.

\noi
But this  procedure is unable to  produce any simplification as soon as $n>2$.
So we face this difficulty that {\em even for free particles\/}, 
the usual form of the equations of motion  fails to permit the elimination 
of superfluous degrees of freedom.

\noi  This point may seem to be academic, because a system of noninteracting 
particles has no  bound state, which renders a 
three-dimensional formulation unnecessary.
But we bear in mind the eventuality of introducing interactions that 
ultimately give rise to bound states. Therefore the possibility of a reduction 
is essential and {\em should survive in the free case\/}.

\medskip
\noi  In this paper we  focus on  three-body systems and
we firstly   consider the case of noninteracting particles.  Let us 
stress that the free system is not considered on its own right, but rather as 
preliminary to the further introduction of mutual interactions.

\medskip
\noi    Since  the KG equations  as they stand   do not permit  a 
factorization of the dependence on relative times,  it is natural to transform 
these equations into an equivalent system such that  two superfluous degrees of 
freedom can be desentangled from the kinematics.

\noi 
An early attempt to carry out this task for an arbitrary number of particles was
 made by Sazdjian \cite {Saz1}\cite{Saz2} fifteen  years ago. 
 Here, however, we shall be concerned with the 3-body case only, and shall
 take advantage of a simplification  that is not possible for $n>3$.

\noi   Our aim is to eliminate two degrees of freedom 
in such a way that the mass-shell constraints reduce to a covariant problem 
with three-dimensional arguments. Ultimate introduction of interactions will 
be briefly sketched at the end.
Of course, the Poincar\'e invariance of kinematics must be preserved and all 
particles should be treated on equal footing (democracy).
 These  conditions are not likely
to select a unique scheme, but if we intend to make it as simple as 
possible, there are not too many choices.

\noi   
We  perform  a rearrangement of the individual 
coordinates (well  known in celestial mechanics) which is adapted to the 
consideration of relative variables.  
We insist on having invertible formulas, which is necessary in order to make 
sure that the new form of the  equations of motion is equivalent to the 
original KG system.

\medskip  
\noi  Section 2  is devoted to an exposition of the notation used and  
of the basic useful equations of relativistic dynamics. In Section 3 we 
 collect known results and perform elementary manipulations. 

\noi
In Section 4, using the "heliocentric  variables",
  we construct in closed form a transformation  of the free-particle system 
and discuss under which conditions this transformation is  invertible.
In Section 5, we briefly indicate how mutual interaction could be  introduced.

\bigskip
\section {Basic equations, notation}

\noi Units are such that $\hbar = 1$ whereas $c$ remains unspecified.

\noi   We start from the KG-system describing $n$ particles 
      in  momentum representation
\begin{equation}
 p_a ^2  \Phi = m_a ^2  c^2  \Phi   \qquad \quad   a,b,c = 1 ...n
                                         \label{weq}            \end{equation}
where $\Phi $ depends on the three four-vectors $p_a ^\alpha$.
 Configuration and momentum variables are mutually conjugate
$ [q_a ^\alpha , p_ {b \beta} ]  =  
i \delta _ {ab} \delta ^\alpha  _\beta , \   $
and so on.          We make use of  the   following notation:
 \begin{equation} Q= {1 \over n}  \sum  q_a ,   \qquad     P = \sum p_a , 
 \qquad    z_{ab} = q_a  -q_b                    \label{2.1} \end{equation}
\begin{equation}     y_a  = {P \over n }  -  p_a    \label{2.2}  \end{equation}
Moreover it is convenient to  define
\begin{equation}   P_{ab} = p_a + p_b    \qquad \qquad
   y_{ab} = \half (p_a -p_b)                 \label{2.3}       \end{equation}
Beware that $z_{ab}$ {\em is not\/} conjugate to $y_ {ab}$.  
We obviously have the following relations
\begin{equation} 
y_a - y_b = -(p_a -p_b) = -2 y_ {ab}    \label{util} \end{equation}
  $$\sum y_a =0 ,  \qquad \quad  2 y_{ab} + y_a  -  y_b  = 0, \qquad
\quad  {\widetilde p }_a  =  -  \ytil _a               $$
The tilde symbol denotes projection orthogonal to $P^\alpha$, in other words
$   {\ytil}_a  =  \Pi    \    y_a   ,
\quad     {\ztil}   _a   =  \Pi   \    z_a   $, with
$   \Pi   =  \delta   -  ( P  \otimes  P) / P ^2    $. Similarly, the "hat" 
symbol refers to the projection orthogonal to $k^\alp$, eigenvalue of $P^\alp$.
For instance    
$${\ytil}_a  ^\sig = y_a  ^ \sig - (y_a  \cdot P / P^2)  \   P^\sig $$
$${\yhat}_a  ^\sig = y_a  ^ \sig - (y_a  \cdot k / k^2)  \   k^\sig $$

\medskip
\noi        {\sl Heliocentric variables}

\noi  
The problem of  "relative times" cannot be easily handled unless we first
 choose a set of independent {\em relative variables}.   For this end, 
  one particle is arbitrarily picked up; 
   let it be particle with label  one.
With respect to particle $1$, the  relative configuration variables are  
defined like in  \cite{llos}.
\begin{equation} z_A =  q_1 - q_A                \label{rel} \end{equation}
where the capital labels $A,B,C$ run only from 2 to $n$.
 From (\ref{2.2}) it follows that  $z_A$ is conjugate to $y_A$.

\noi      Let us now specialize to  three-body systems;  we can write 
\begin{equation} y_{12} = y_2 + \half y_3   \qquad  \qquad
   y_{13} = y_3 + \half y_2    \label{difp}   \end{equation}
\begin{equation}  z_{12} + z_{23} + z _ {31} = 0  \label{zed} \end{equation}
Notice that eqs (\ref{weq}-\ref{util}) hold true for any $n$,  
whereas   (\ref{difp})(\ref{zed}) are  valid for $n=3$ only.
  It is clear that                      
 $Q, z_2 , z_3 $ are independent configuration variables.
In the same way $P, y_2 ,y_3 $ are independent momentum variables,
 canonically conjugate to them. 
 We can use the  set of canonical variables  
$Q, z_2, z_ 3 , P , y_2 , y_3 $ 
in place of         $q_1, q_2, q_3, p_1 , p_2 , p_3 $, this change is trivial.

\noi  In this "heliocentric"  formulation, democracy among the three 
particles is of course not kept manifest but can be checked at various stages 
of the development.
A similar re-arrangement, showing up two relative momenta, is of current use in 
(Newtonian) celestial mechanics.

\medskip
\noi     Among the quantities $P_{ab}$
we shall more specially need to evaluate   $P_{12},  P_{13}$.
 They are given by
\begin{equation}               P_{12}   =  {2 \over 3}  P  + y_3 \qquad   \qquad
 P_{13}    =  {2 \over 3}  P  + y_2      \label{sub}      \end{equation} 
We shall also need the canonical expression of $y_{12}, y_{13} $,
 given by (\ref{difp}).

\noi  It will be  convenient to replace  eqs (\ref{weq}) by their sum and their 
differences; to this end we  define 
$$ \nu _A = \half  (m_1 ^2 - m_A ^2)                          $$
so the equal-mass case is characterized by the vanishing of both
 $\nu _1 , \nu _2$.

\bigskip
\section {Equations of motion.}

\noi Equations (\ref{weq}) can obviously be written 
\begin{equation}  c^2   \sum m_a ^2   \   \Phi  =  
 \sum p_a ^2  \     \Phi      \label{V} \end{equation}
\begin{equation}  (m_a ^2 - m_b ^2) c^2   \Phi   = 
 (p_a^2  -p_b ^2)  \Phi   \label{Vbis} \end{equation}
Notice that, according to notation (\ref{2.3}) 
 \begin{equation}   \half (p_a ^2 - p_b ^2) =
 y_{ab}  \cdot P_{ab}      \label{difH} \end{equation} 

\medskip

\noi  In equation (\ref{V}), let us use the  identity
\begin{equation} n \sum_1  ^n   p^2       \equiv  
P^2 + \sum _{a<b} (p_a - p_b )^2     \label{oldiden}      \end{equation}
valid for any sum of $n$ squares in a commutative algebra.
We obtain
\begin{equation}      3 \sum  m^2  c^2   \    \Phi  =  P^2  \   \Phi +
 \sum _{a<b}   (p_a -p_b )^2    \   \Phi      \label{6somH}   \end{equation}
In terms of the relative variables (see eq. (\ref{difp}))  we have  another 
 identity {\em specific of the three-body problem\/}
\begin{equation}     \sum _{a<b}   (p_a -p_b )^2   \equiv
6 (y _2 ^2  + y_3 ^2  +  y_2 \cdot  y_3 )       \label{newiden}   \end{equation}
Now in the r.h.s. of (\ref{newiden}) we separate time from space according 
to the direction
 of $P$, and insert the result  into  (\ref{6somH}). We get 
\begin{equation}   \sum _{a<b}   (p_a -p_b )^2   = 
D  + 6 P^2 \Xi     \label{som} \end{equation}
where 
\begin{equation} 
 D = 6 (\ytil  _2 ^2  + \ytil _3 ^2  +  \ytil _2 \cdot  \ytil _3 )
                                               \label{defD}  \end{equation}
\begin{equation} \Xi =
 (P^2) ^{-2} [ ( y_2 \cdot  P )^2  +  (y _3 \cdot P ) ^2  +
 (y _2 \cdot  P )( y_3 \cdot P )  ]       \label{dejaXi}        \end{equation}
Thus the sum of eqs (\ref{weq}) is 
\begin{equation} (3 \sum m^2  c^2  - P^2 ) \   \Phi = 
       (D  + 6  P ^2 \Xi  ) \    \Phi           \label{sumeq}    \end{equation}
 The remaining combinations of (\ref{weq}) can easily be written as the 
"difference equations"
 \begin{equation} y_{12}  \cdot P _{12} \   \Phi =
 \nu _2 c^2  \     \Phi,      \qquad \qquad      y_{13}  \cdot P _{13}
  \   \Phi = \nu _3  c^2   \  \Phi    \label{weqnu}      \end{equation}
Now it is natural to require that $\Phi $ is also eigenstate of the total 
momentum, say
\begin{equation}
  P^\alpha   \Phi  = k ^\alpha \Phi       \label{P}           \end{equation}
for some timelike constant vector $k$. 
But (in contrast to what happens in the two-body case)  this procedure is 
unable of   getting rid of  the relative energies  
$ \disp   c  y_A \cdot  k  /  \sqrt {k^2}$  conjugate to the relative times
$\disp   c^{-1} \    z_A \cdot k /   \sqrt {k^2}    $.

\medskip 
\noi  Nevertheless, we can look for a new set of canonical variables; 
if these variables are suitably choosen,     equations  (\ref{weqnu})
 may  after all result in the elimination of two degrees of freedom.

\medskip

\bigskip
\section {Alternative Formulation of the  Free Motion}

\subsection {Transformations  in momentum space}

\noi    We shall construct a new representation of the KG system. It will 
involve a new set of operators 
$q'_a , p'_b$  satisfying the canonical commutation relations.
Let them be rearranged  as  $P', z'_A , y'_B $ by formulas similar to 
(\ref{2.1})(\ref{2.2})(\ref{rel}).
In particular $ \sum y'_a $ vanishes and $P' = \sum p'_a \  $ but we must 
require that $P'=P$ in    order to preserve translation invariance.  Thus
$$ y'_a  = {P \over n }  -  p'_a   \qquad  \qquad
 y'_{ab} =  \half (p'_a -p'_b)    $$ 
$$ y'_a - y'_b = -(p'_a -p'_b) = -2 y'_ {ab}    $$
Naturally     $\sum y'_a \equiv 0$. 
Notice  for $ \disp  \sum _1 ^3  {p'}^2 $ 
and for          $  \displaystyle 
\sum _{a<b}   (p'_a -p'_b )^2  $  identities  similar to 
(\ref{oldiden}) and (\ref{newiden}).      
 We obtain
\begin{equation}y'_{12} = y'_2 + \half y'_3    \qquad
 \qquad  y'_{13} = y'_3 + \half y'_2         \label{difp'}   \end{equation}
\begin{equation}  y'_2 =  {4\over 3}  y'_{12} - {2 \over 3}   y'_{13} 
\qquad    \qquad
    y'_3 =  {4\over 3}  y'_{13} - {2 \over 3}   y'_{12}
                                               \label{fidp'}   \end{equation}
Define
\begin{equation} Q' = {1\over 3}  \sum q'_a       \qquad \qquad 
   z'_A =  q'_1 - q'_A                             \label{rel'}  \end{equation}
It is clear that $Q', z'_2 , z'_3 , P, y'_2 ,y'_3 $ 
must  be  independent variables, $y'_A $ conjugate to $z'_A$, etc. 

\noi 
A  transformation in momentum space will be enough to induce  the suitable  
transformation among operators. 
 In fact we are going to construct the quantum analog of a {\em point 
transformation} in momentum space (see Appendix 1).

\noi       Let us start with a wave function  $\Phi (p_1, p_2, p_3) $.
Perform    a  change  in the space of its arguments
$$  p_a     \mapsto     p'_b    $$
or equivalently 
$ P \mapsto  P' = P$  and $  y_A  \mapsto   y'_A   $.

\noi    Instead of the old configuration variables
$  \disp    \    z  =  i  {\partial  \over \partial y},     \qquad   \quad
   Q   =   i          {\partial  \over \partial P}  \   $,  
we shall now consider
$$  z'  =  i  {\partial  \over \partial y'}        \qquad  \quad
  Q'   =   i          {\partial  \over \partial P'}              $$
Since  $ \partial P  /  \partial y'  = 0 $ and
 $ \partial P  / \partial P' =  \delta $,
 the transformation formulas are  as follows       
   \begin{equation} {z'}_A ^\alpha =
  {\partial y_B ^\sigma  \over  \partial   y'_{A \alpha}}  \    z _{B \sigma}
                                               \label{confi1}   \end{equation}
\begin{equation} {Q'} ^\alpha  =      Q ^\alpha  +
  {\partial y_A  ^\mu   \over  \partial   P'_\alpha}  \    z _{A \mu}  
                                                 \label{confi2}  \end{equation}
with  summation also over (repeated) capital indices.
  In these  formulas it is clear   
that the transformation of momenta must be invertible.

\noi     Beware that $Q'$ may not coincide with $Q$ because  of  
$  {\partial y  / \partial   P'}  $.  In addition, we observe that the new 
relative coordinates actually  mix the  old ones. 
However, we shall prove later (Section 5) that this difficulty  disappears 
 in the large-total-mass limit.

\bigskip
\noi  It is in order to stress that finding the desired  
transformation amounts to solve a problem in the framework of $c$-numbers. 
The question of inverting formulas, discussed below,  is nothing but a
 nonlinear  problem concerning the arguments of the wave function.
Since it is specified that we are dealing with momentum representation, we 
shall use without confusion the same symbols for the arguments of the wave 
function and the {\em multiplicative operators\/} they define.

\medskip
\noi   For a better understanding of the mathematical structure,
 it is perhaps relevant to notice that         $Q'$ and $z'_A$ are  
"formally hermitian" in this sense that they are symmetric operators in  
$$ {\cal L '}^2 ({\bf R} ^{12}) =
 L^2 ({\bf R} ^{12}, \    d^4 P \  d^4 y'_2 \   d^4 y'_3)$$
 whereas   $Q$ and $z_B$ are symmetric operators  in  
$$   {\cal L}^2 ({\bf R} ^{12}) =
L^2 ({\bf R} ^{12}, \    d^4 P \  d^4 y_2 \   d^4 y_3) $$
In contrast the momenta are symmetric operators in both senses.
For mathematical convenience we shall  work with a new wave function 
$ \disp  \Psi =  |J| ^\half  \Phi$,  where $J$ is the Jacobian
 $ \disp   J =  {D (p_1, p_2, p_3)  \over D (p'_1 , p'_2 , p'_3) } $
always finite and nonvanishing insofar as   our transformation is invertible.
Indeed multiplication by   $   |J| ^\half$  maps  $ {\cal L}^2 $  onto
$ {\cal L'}^2 $, so  $\Phi$ (resp. $\Psi$ ) belongs to the rigged-Hilbert
 space constructed  by taking 
 ${\cal L}^2 $  (resp.   ${\cal L'}^2 $) as Hilbert space.
Although   ${\cal L}^2 $  (resp.   ${\cal L'}^2 $) has no direct physical 
meaning, it allows for representing the Poincar\'e algebra and gives a 
rigorous status to the operators involved in the wave equations.

\noi  
Since $p_a$ are multiplicative operators, they commute with $J$, so the mass-
shell constraints can be written either as (\ref{weq}) for $\Phi$ or
 equivalently in the form
$ \disp  (p_a ^2  - m_a ^2  c^2 )          \Psi = 0$,
 with each $p_a$ expressed in terms of $p'_b$.

\noi   
In momentum space, the Lorentz group is characterized by this  property that 
it leaves all the products  $p_a \cdot p_b$ unchnged.

\noi {\em Provided all the   $p'_a \cdot p'_b$ can be expressed as  functions 
of         $p_c \cdot p_d$  and vice-versa\/}, the same realization of the 
Lorentz group can be as well characterized by invariance of all the scalar
 products  $p_c \cdot p_d$.
In such a situation, although $M' = \sum q' \wedge p' $ may be distinct from
   $M = \sum q \wedge p $,  their components span the same Lie algebra.
Moreover $J$ being conserved by rotations, it follows that $M$ and $M'$ are 
both symmetric in ${\cal L}^2$  and also in ${\cal L'}^2 $.

\bigskip
\noi  Till now we have considered a large class of transformations, 
characterized by equations  (\ref{confi1})(\ref{confi2}); 
the  classical (non-quantum) limit  of such formulas would define 
 {\em point transformations in momentum space}.
 
\noi
We now specialize to a transformation which allows for eliminating  the 
 superfluous degrees of freedom.  
All we need is an invertible  transformation such that 
\begin{equation} P_{12} \cdot  y_{12}  = 
  P \cdot   y'_{12}  \qquad \  P_{13} \cdot  y_{13}  = 
  P \cdot   y'_{13}  \label{basic23}       \end{equation}
Indeed, if these relations are satisfied, (\ref{weqnu}) takes on the form 
\begin{equation} y'_{1A} \cdot P  \  \Psi  
=  \nu _A  c^2  \   \Psi                    \label{basic}      \end{equation}
Then  according to (\ref{fidp'}) the "difference equations" are 
\begin{equation} y'_2 \cdot P   \   \Psi =
  (  {4\over3} \nu _2  -   
{2 \over 3} \nu _3 )   c^2  \      \Psi       \label{nu2}      \end{equation}
\begin{equation} y'_3 \cdot P   \   \Psi =  (  {4\over3} \nu _3  -
   {2 \over 3} \nu _2 )  c^2  \   \Psi       \label{nu3}        \end{equation}
With help of equation (\ref{P}) we obtain 
$$  \Psi   =   \qquad  \qquad     \qquad                 $$
\begin{equation}   \delta (P^\alpha - k^\alpha)  \
\delta ( y'_2 \cdot k -   {4\over3} \nu _2 c^2  
 -   {2 \over 3} \nu _3   c^2        ) \
\delta ( y'_3 \cdot k -   {4\over3} \nu _3 c^2 
 -   {2 \over 3} \nu _2     c^2    )
   \    \psi                                     \label{psi}  \end{equation}
where $\psi $ depends  on $y'_2 ,  y'_3$ only through their orthogonal
projections onto the three-plane orthogonal to $k$.
 One remains with the problem of determining 
a  reduced (or {\em internal}) 
wave function  $\psi$  which has no more arguments than the wave 
function of a nonrelativistic problem. 
The dependence of the wave function on
  $ y' _A  \cdot  k$ is now factorized out. 
        
\bigskip
\noi
For simplicity we complete our  transformation law by imposing that 
  the space projections  of $y_2, y_3$ (with respect to the rest frame) remain 
unchanged, say
\begin{equation} \ytil ' _A =  \ytil _A   \label{tild=}        \end{equation}
and just transform  their  time components in the  way dictated by  eqs
(\ref{basic23}). 
This choice obviously  preserves Lorentz invariance; we shall prove below that 
it does not destroy the democracy among  particles.

\bigskip   
\noi In view of equation (\ref{tild=}), and taking into account the identity  
\begin{equation} y'_ {A \alpha}  \equiv
 {y'_A \cdot P \over  P^2} P_\alpha  +  \ytil ' _{A \alpha} 
                                              \label{decompo}    \end{equation}
it is clear that our change of variables is essentially  determined by
(\ref{basic23}). As they stand, these formulas implicitly define 
$ y'_2 \cdot P $ and $y'_3 \cdot P$ in terms of the old variables; but we 
still have to solve (\ref{basic23}) for  $ y'_2 \cdot P $ and $y'_3 \cdot P$
in order to exhibit the transformation in closed form. 
   
\noi   According to (\ref{sub}) and (\ref{difp}) the left-hand sides of 
conditions   (\ref{basic23})                   are as follows:
\begin{equation} P_{12} \cdot  y_{12}  =
{2\over 3} y_2 \cdot P + y_2 \cdot y_3 + {1 \over 3} y_3 \cdot P 
+ \half  y_3 ^2                \label{lhs2}       \end{equation}              
\begin{equation} P_{13} \cdot  y_{13}  =
{2\over 3} y_3 \cdot P + y_2 \cdot y_3 + {1 \over 3} y_2 \cdot P 
+ \half  y_2 ^2           \label{lhs3}            \end{equation}              
For  the  right-hand sides, eqs (\ref{difp'}) yield
\begin{equation}  y'_{12} \cdot  P  = 
(y'_2 + \half y'_3) \cdot  P   \label{rhs2}        \end{equation}
\begin{equation}  y'_{13} \cdot  P  =
 (y'_3 + \half y'_2) \cdot  P   \label{rhs3}        \end{equation} 
Therefore the requirement that (\ref{basic23}) are satisfied can be
 expressed as the linear system
\begin{equation}
   {2\over 3} y_2 \cdot P + y_2 \cdot y_3 + {1 \over 3} y_3 \cdot P 
+ \half  y_3 ^2   =
      (y'_2 + \half y'_3) \cdot  P          \label{sys2}    \end{equation} 
\begin{equation} 
   {2\over 3} y_3 \cdot P + y_2 \cdot y_3 + {1 \over 3} y_2 \cdot P 
+ \half  y_2 ^2    = 
  (y'_3 + \half y'_2) \cdot  P            \label{sys3}  \end{equation}
to be solved for $y'_2 \cdot P $ and  $   y'_3 \cdot P $.
The outcome  of system (\ref{sys2})(\ref{sys3}) is 
\begin{equation} y'_2 \cdot P =
 {2\over 3} (  y_2 \cdot P   +y_2 \cdot y_3 + y_3 ^2)   
    - {1 \over 3}  y_2 ^2                    \label{sol2}  \end{equation}
\begin{equation} y'_3 \cdot P =
  {2 \over 3}  (y_3 \cdot P   + y_2 \cdot y_3   + y_2 ^2 )
       - {1 \over 3}  y_3 ^2                   \label{sol3}  \end{equation}
whereto we insert the decomposition (\ref{decompo}). 
This substitution, together with (\ref{tild=}), determines 
in closed form the transformation of momenta.   But it remains to be checked 
that this transformation is invertible.
                                                                             
\bigskip
\noi  {\sl Translation invariance}
 was ensured from the outset by assuming that $P'=P$.

\noi {\sl Lorentz invariance}
 is preserved because all the quadratic scalar quantities 
formed with the vectors $P, y_2 , y_3 $  are scalar invariant under spacetime 
rotations.

\noi
{\sl  Democracy}  between  particles is not  manifest in the 
heliocentric notation.
Nevertheless it is not difficult to check that our way of transforming 
momentum variables treats all three particles on the same footing. Indeed we 
first observe that (\ref{tild=}) entails
 $ {\widetilde y} '_1 =  {\widetilde y}_1 $,  
which amounts to finally write
$ \disp  {\widetilde y}'_a   = \widetilde y_a $ for the three particles.
 Then using (\ref{difH}) and (\ref{2.3}) 
we realize that (\ref{basic23}) automatically imply a third relation 
$ P_{23} \cdot y_{23}  =  P \cdot y' _{23} $.

\bigskip
\subsection {Inversion of formulas}

\noi
 Now that all components of the new  momenta $p'_a$  are determined 
  we can (in principle) evaluate 
the configuration variables through formulas (\ref{confi1})(\ref{confi2}).
It is essential to realize
 that our transformation of the momenta among themselves must be invertible:
  if it were not, the transformation would not be canonical and the 
new form given to the  wave equations  would not be  equivalent with the KG 
system.

\noi  
Formula (\ref{confi1}) can  be written in closed form provided we are able to 
carry out this inversion.
We are thus faced with the problem of mapping the new momenta  back onto 
the old ones, which amounts to solve  the system (\ref{sys2})(\ref{sys3})
now   for  the unknown
  $ y_2 \cdot P , \    y_3 \cdot  P $  in terms of   
$  y'_2 \cdot P ,  \   y'_3 \cdot P $,
assuming this time that the latter are given and taking 
 (\ref{tild=}) into account.

\medskip   \noi   {\sl Positive-energy condition}

\noi
The domain where (\ref{sys2})(\ref{sys3}) must be inverted can be
 limited  to the 
{\sl positive-energy  sector}. So we  require not only that $P$ is timelike and
 future oriented,  but also   that 
{\sl every  vector  $p_a$  is timelike and points
 toward the future\/}, which entails  $ P \cdot  p_a  > 0 $  and 
$ p_a  \cdot  p_b  > 0 $.

\medskip
\noi 
 At this stage it is convenient to introduce the  dimensionless quantities
\begin{equation} \xi ={ y_2 \cdot P  \over P^2}, \qquad \qquad   \quad 
 \eta = { y_3 \cdot P    \over P^2}            \label{defxi}   \end{equation}
thus (\ref{dejaXi}) becomes
\begin{equation} \Xi  =
  \xi ^2 + \eta ^2 + \eta \cdot \xi     \label{defXi}        \end{equation}
The positive-energy condition above implies limitations for 
$\xi$  and $ \eta$. Indeed we first derive  from (\ref{2.2})  
\begin{equation} \xi = {1\over3} - {P \cdot p_2  \over P^2}      \qquad   \quad
 \eta = {1\over3} - {P \cdot p_3  \over P^2}    \label{posit}    \end{equation}
 From positivity of $P \cdot p_A$ we get
\beq  \xi  < {1 \over 3}, 
 \qquad   \quad      \eta  < {1 \over 3}     \label{ineq1}   \eeq 
\noi
  On the other hand we have  $P\cdot p_1 = P^2 - P \cdot p_2  - P \cdot p_3  $. 
According to  (\ref{posit})  this identity reads
$   \disp   P\cdot p_1 /  P^2 = {1\over 3} + \xi +\eta     $  and 
 this expression also must be  positive.  We end up with
\begin{equation}  -{1 \over 3}   <   \xi + \eta      
                                        \label{ineq2}    \end{equation}

\medskip \noi
With these limitations in mind, we now turn to the inversion of system 
(\ref{sys2})(\ref{sys3}).
In view of   the  identities
 $$ y_A \cdot y_B                   \equiv
  {(y_A \cdot P)( y_B \cdot P) \over  P^2 }+ \ytil _A \cdot \ytil _B    $$
we can write  
$$y_2 \cdot y_3 = P^2 \xi \eta  + \ytil _2 \cdot \ytil _3     $$
$$ y_2 ^2 =  P^2 {\xi  ^2 } + \ytil _2 ^2  ,   \qquad     \qquad
   y_3 ^2 =  P^2  {\eta ^2 } + \ytil _3 ^2   $$
Insert these formulas 
 into (\ref{lhs2})(\ref{lhs3}), and write (\ref{basic23}). We get 
\begin{equation}  {2 \over 3} \xi + {1 \over 3} \eta + { \xi \eta }  + 
{\eta ^2 \over 2 } + {\ytil _2 \cdot \ytil _3  \over P^2}   +
 \half {\ytil _3  ^2  \over P^2}
={ y'_{12} \cdot P   \over P^2}               \label{quadr2} \end{equation}
\begin{equation}     {2 \over 3} \eta + {1 \over 3} \xi + { \xi \eta }  +
{\xi ^2 \over 2} + {\ytil _2 \cdot \ytil _3 \over P ^2}
+ \half  {\ytil _2 ^2 \over P^2} =
{ y'_{13} \cdot P  \over P ^2}         \label{quadr3}    \end{equation}
Because of
 (\ref{tild=}) all quantities of the form $ \ytil _A \cdot \ytil _B$ are 
already   known. 

\noi
     The above  system (\ref{quadr2})(\ref{quadr3}) is quadratic in the unknown 
quantities  $\xi ,  \eta$.
Define dimensionless quantities $u, v$ through the formulas
\begin{equation}  P^2 u=   y'_{12} \cdot P  - 
 ( \ytil _2 \cdot \ytil _3  + \half {\ytil} _3 ^2 ),      \qquad \qquad
P^2 v=  
y'_{13} \cdot P  -  ( \ytil _2 \cdot \ytil _3  + \half  {\ytil} _2 ^2)
                                                   \label{uv} \end{equation}
They are regarded  as functions of the  new momenta,
 since $P$ and $ \ytil _A $ coincide with  $P'$ and $\ytil '_A $ respectively.
Inserting (\ref{difp'})  into (\ref{uv})  yields
\beq  P^2 u =    y'_2  \cdot  P  +  \half y'_3 \cdot  P 
-  { \ytil }' _2  \cdot {\ytil}'_3 - \half  ({\ytil }'_3)^2              
                   \label{ubis}          \eeq           
\beq  P^2 v =    y'_3  \cdot  P  +  \half y'_2 \cdot  P 
-  { \ytil }' _2  \cdot {\ytil}'_3 - \half  ({\ytil }'_2)^2              
                   \label{vbis}          \eeq
The system  (\ref{quadr2})(\ref{quadr3})    becomes
\begin{equation}  {2 \over 3} \xi + {1 \over 3} \eta + { \xi \eta }  + 
{\eta ^2  \over 2}   =  u                \label{linxi} \end{equation}
\begin{equation}     {2 \over 3} \eta + {1 \over 3} \xi + { \xi \eta }  +
{\xi ^2 \over 2 }  =     v                 \label{lineta}    \end{equation}
to be solved for $\xi, \  \eta  \   $ with $u, v$ as in (\ref{uv}).   Setting 
$$ 3 (u+v) = \sigma, \qquad  \qquad     3(u-v) = \epsilon            $$ 
system (\ref{linxi})(\ref{lineta}) can be cast into the form
\begin{equation}
 \xi + \eta + 2 \eta \xi + \half (\xi ^2 + \eta ^2 ) = {\sig \over 3}  
                                                \label{sig}  \end{equation}
\begin{equation} {1 \over 3} \xi  - {1 \over 3} \eta 
+ \half (\eta ^2 -\xi ^2 ) = {\epsilon \over 3}     \label{eps}   \end{equation}
It is convenient to  define 
$   \    X  = \xi + \eta  ,   \quad        Y = \xi - \eta $.
When inserted into   (\ref{dejaXi}) this change of variables produces 
\begin{equation}
 \Xi = {3 \over 4} X^2 + {1\over 4} Y ^2      \label{Xinew}  \end{equation}
System (\ref{sig})(\ref{eps}) becomes
\begin{equation} {3\over 4} X^2 + X  -{1 \over 4} Y ^2 =    {\sig \over 3}    
\label{sig'}  \end{equation}
\begin{equation} 
    2 Y - 3 XY    =  2 \epsilon             \label{eps'}   \end{equation}
The positive-energy conditions  (\ref{ineq1})(\ref{ineq2}) demand 
  that $X$ belongs to the open interval 
$\disp   (-{1\over3} , {2\over 3} ) $   and  also that  $Y>-1$,  which in turn  
require that $\disp   X <  {2 \over 3} (1 +  \epsilon)$.

\medskip

\noi  When   $\epsilon = 0$  a couple of  obvious  solutions is  given by
$X= \disp   {2 \over 3}$ (whatever is $\sig$) which corresponds to 
$  Y  =   \pm 2 \sqrt{1 -  \sig   /  3  }     $,   
but this possibility is ruled out by  (\ref{ineq1}).  
Other solutions are given by 
 $Y= 0$  hence  
\begin{equation}  X  = X ^\pm            =       
 {2 \over 3} ( -1  \pm \sqrt {1 + \sig})    \label{Xpm}   \end{equation}       
but  the solution $X^-$ is excluded in view of condition (\ref{ineq2}).

\medskip
\noi    We now turn to the general case. The possibility that   strictly
$X = 2/3$   being discarded,         we now solve (\ref{eps'}) 
\begin{equation} Y =  {2\epsilon  \over  2 - 3 X}  \label{igrec}  \end{equation}
and bring the result into   (\ref{sig'}). Hence
 a  4th-degree polynomial equation to solve for $X$,
\begin{equation}  (2-3X)^2 ({3 \over 4}X^2 + X - {\sig \over 3})
 =  \epsilon ^2      \label{deg4} \end{equation}
{\sl Graphic analysis}

\noi   In principle such equation can be explicitly solved by radicals. But 
a graphic analysis gives a better understanding.
Solving (\ref{deg4})  amounts to  discuss how, in the $X,Z$ plane,
 the parametrized curves
 $\disp   
Z =  R_\sig (X) =    (2-3X)^2 ({3 \over 4}X^2 + X - {\sig \over 3})    $ are 
intersected by a straight line with trivial equation  $Z= \epsilon ^2$.
See Appendix 2.
                 
\noi      The outcome of graphic analysis is:

\noi      {\bf  Proposition I}.

\noi        {\sl     Provided $\disp  -{3\over 4} < \sig  <  \half $ and
 $\epsilon$  is  taken in the open interval 
$\disp  (-\half ,   {1 \over 2} )$,
 among the real solutions  of the system (\ref{sig'})(\ref{eps'})
 there exists a unique one, $ X, Y $     such that  
  $  \disp   X \in (-{1\over 3} , {2 \over 3})$   and such that 
$X$ reduces to $X^+$ when $\epsilon$  vanishes.\/}

 \noi  Moreover we observe that

 \noi  {\sl   $\disp  X - {2 \over 3}  \epsilon$  remains bounded
 by  $\disp  {2 \over 3}$,  ensuring that  $Y >  -1$ as required among the
 positive-energy conditions\/}.

\medskip  \noi
The expression $   X = S(\sig, \epsilon)$  for this solution could be written 
in closed form, but is very complicated, except naturally for vanishing 
$\epsilon$ where it is just given by $X^+$.
For applications, we have better to  use a development in powers of
 $\epsilon ^2$,  say 
  \begin{equation}X  =   S(\sig , \epsilon) =
  X^+  + \epsilon^2  X_{(1)}
+ \epsilon ^4  X_{(2)} .....+\epsilon ^{2p} X_{(p)} 
 +  .... \label{dvlop}   \end{equation} 
All coefficients $X_{(p)}$ are derived from (\ref{deg4}) and depend on $\sig$.
 We find for instance
 $\disp  X_{(1)} = {4 \over 3 (X^+ - X ^-) (2- 3 X^+)^2 }   $.
Note that    
\beq      S =  {\sig \over 3}  +  O ( \sig ^2 , \epsilon ^2 ,
           \epsilon  \sigma )          \label{dvlopS}                \eeq

For the sake of a physical interpretation, investigating the behavior of our 
formulas at large $P^2$ is of interest. Equations (\ref{ubis})(\ref{vbis})
show that, considered as functions of the independent variables 
${y'_A}^\alp ,  P^\beta$, all the quantities $u, v, \sig ,\epsilon, X, Y $
are of the order of $ 1/|P|$. We simply have
\beq \xi = 2u -v + O(1/P^2) , \qquad   \quad
      \eta = 2v-u + O(1/P^2)                       \label{etaxilin}  \eeq
\medskip 
 \noi
 Proposition (I) stated  above ensures  that the transformation from the old 
momenta to the new ones is safely invertible in an open set of values given to 
the couple $\sig , \epsilon$.
As these quantities are first integrals for free particles, 
 their limitation to an interval  defines a sector which is invariant by the 
motion.  Characterization of this sector in terms of physical quantities will 
be discussed in the next Section.

 \noi Remark:

\noi  Infinitely many other domains ensuring  a unique solution to 
(\ref{sig'})(\ref{eps'})   could be 
exhibited. But we can enlarge the interval for $\epsilon$ only at the price of 
shrinking the one for $\sig$.

\bigskip
\noi 
\subsection {Physical conditions}

\noi      In view of eqs. (\ref{basic}), the wave function includes a factor 
$ \delta  (y'_{12} \cdot P -  c^2  \nu _2) 
               \delta   (y'_{13} \cdot P -  c^2   \nu _3)      $.
The relevant domain for the arguments of $\Psi$ is thus limited by the 
constraints   $y'_{1A} \cdot  P =  c^2  \nu _A $,
 where the masses are given from the outset.
 On the mass shell, we can replace $y'_{1A} \cdot P $ by $ c^2 \nu _A$ in 
(\ref{uv}) or in the definitions of $\sig, \epsilon$.
   
\medskip
\noi  The particular case where $\epsilon$ vanishes is  interesting because it 
arises when the  particles are mutually at rest, provided $m_2 = m_3$,  which 
includes the  special case where all masses are equal.
Moreover $\epsilon$  remains  small insofar as  $\nu_2, \   \nu_3$ and the 
{\em velocities\/} are not too large.

\noi
For simplicity,  let  us focus on the assumption that    $\nu_2, \   \nu_3$ 
 are small enough. In order to keep some contact with 
nonrelativistic mechanics, our scheme must encompass the case $\epsilon=0$;
thus the  solution which reduces to $X^-$ for vanishing $\epsilon$ is excluded.
Since the transformation of momenta must  be one-to-one, 
we are also obliged to discard the solutions which reduce  to the  fixed point  
for vanishing $\epsilon$.

\noi  
 Finally we have no other choice than the solution given by 
$X= S(\sig, \epsilon)$.

\medskip
   Let us now  discuss in  more details  how we can manage, by simple physical 
requirements,     to keep $\sig, \    \epsilon$
 within  admissible values  allowing to apply Proposition I.

\noi      From (\ref{uv}) we obtain
\beq P^2 \    {\sig \over 3}  = (\nu _2 + \nu _3 ) c^2  -
(2 \ytil _2 \cdot \ytil _3 + \half {\ytil}_2 ^2  + \half {\ytil} _3 ^2 )
                                                         \label{64.1}      \eeq
\beq P ^2\     {\epsilon \over 3} = (\nu _2 - \nu _3) c^2   +
\half (   {\ytil} ^2 _2   -  { \ytil} ^2 _3) 
                                                     \label{64.2}          \eeq
in other words
\beq    \sig = \sig _0   - {3 \over P^2} \   
(2 \ytil _2  \cdot  \ytil_3   +  \half {\ytil_2} ^2   + \half {\ytil _3}^2  )  
                                                             \label{64.I}  \eeq
\beq     \epsilon  = \epsilon _0  +   {3 \over  2P ^2}\ 
                  ( {\ytil _2}^2   -   {\ytil _3 }^2  )    
                                                          \label{64.II}    \eeq
setting  
\beq   \sig _0  =  3 (\nu _2 + \nu _3 )  c^2  / P^2 ,   \qquad   \qquad
 \epsilon _0   = 3 (\nu _2 - \nu _3 ) c^2   / P^2    \label{defsigzero}   \eeq

\noi    
In the domain where the arguments of $\Psi$ vary,  we can for instance impose
 a  democratic  condition 
\begin{equation} 
   |{\ytil} _a  ^2 |  <      
        {P^2  \over  24}      \label {slow}   \end{equation}
We remember that     $  \widetilde p_a =- \ytil _a$,
 thus condition (\ref{slow}) is a   statement about individual momenta. 
From (\ref{slow}) it follows that
\beq  |\sig |   \leq    |\sig _0 |   + {3 \over P ^2}\  
 |2 \ytil _2  \cdot  \ytil _3   +  \half {\ytil_2} ^2 
                                 + \half {\ytil _3}^2    | 
                                                   \label{64.III}   \eeq
 \beq   |\epsilon |  \leq   |\epsilon _0 |   + 
{3 \over  2 P ^2} \      |{\ytil _2} ^2    -  { \ytil _3} ^2  |
                                                     \label{64.IV}   \eeq
  Since every $\ytil$ is spacelike,
 $|\ytil_A  \cdot \ytil_B|       \leq |\ytil_A  | \   |\ytil_B|$.
 Hence  (\ref{slow}) implies
$$ |2 \ytil _2 \cdot \ytil _3 + \half {\ytil}_2 ^2  + \half {\ytil} _3 ^2|
\leq    
{ P ^2  \over 8}        $$
$$ | \half {\ytil} ^2 _2   -  \half { \ytil} ^2 _3 |    \leq   
                                 {P^2  \over 24}        $$ 
Therefore   
\beq   |{\sig } -       \sig _0|       <    
                             {3 \over 8}   \label{64.V}      \eeq
\beq     |{\epsilon } -   \epsilon _0| <  
                             {1 \over 8}   \label{64.VI}     \eeq
\noi  Now, provided that 
\beq   |\sig _0 |   \leq    {1 \over 8} ,  \qquad  \qquad
   |\epsilon _0 |   \leq   { 3\over 8}    \label{64.VIII}       \eeq
it stems from (\ref{64.V})(\ref{64.VI}) that 
$\sig$ and $\epsilon$  remain  within the interval  $(- \half ,  \half) $.
In order to realize this situation we are led to restrict the squared-mass 
differences by the condition  (\ref{64.VIII}).
\noi Then, condition (\ref{slow})  permits to apply Proposition I.

\medskip   \noi
Untill now, we have proposed  condition (\ref{slow}) which involves not only 
the relative momenta but also $P^2$.  Since we consider the positive-energy 
sector of free particles, it is clear that
$$ P^2 >  \sum  p_a ^2 =  \sum  m_a  ^2   c^2    $$
  For the sake of a simple kinematic interpretation,
we have better to replace    (\ref{slow}) by the stronger condition
\begin{equation} 
 |{\widetilde p} _a ^2 |   <  {1 \over 24}    \sum   m_a ^2   c^2
                             \label{slowbis}    \end{equation}
which  is just a little more restrictive 
 and offers  the advantage of  involving only masses and spatial velocities.

Similarly, in view of (\ref{defsigzero}) it is clear that, in order to 
fullfill (\ref{64.VIII}), it is sufficient to demand  
\begin{equation}
    |\nu_2 +   \nu _3 |  <  \sum {m_a ^2 \over 24}    \qquad  \qquad
    |\nu_2 -   \nu _3 |  <  \sum {m_a ^2 \over 8}
                               \label{difmas}             \end{equation}
This approach is well-suited for the equal-mass case  and remain useful when 
the mass differences are not too large.

\noi  {\sl  Example. Two equal masses}.

\noi  Assume that 
$m_A  = \rho m_1 $, hence  $\disp  \sum m_a ^2 = (1+2\rho ^2)  m_1 ^2   $.
We find  that (\ref{difmas}) is satisfied  provided the square-mass ratio 
satisfies
$ \disp    {23\over 26}   <  \rho ^2  <    {25\over 22} $.

\noi It is clear that      (\ref{slowbis})    is a condition on the 
three-dimensional velocities  with respect to the rest frame.
Although it puts a bound on these quantities, it still leaves room for a large 
class of relativistic motions.

\medskip 
\noi  {\sl    Example. Three equal masses}.
In  the equal-mass case, $ m_a = m$, thus both  $\nu _A$ vanish. 
We are sure that $\sig,  \epsilon$ 
belong to the safety interval if we demand that  
\begin{equation} 
| {\widetilde p}_a  ^2 | < {m ^2  c^2  \over 8} 
                                         \label{sslow} \end{equation} 
Indeed positivity entails that  $3 \disp  m ^2  c^2   \leq  P ^2$. 

\medskip      
\noi  Now what does mean (\ref{sslow}) in terms of (Newtonian)  velocities ?
In the rest frame, for all indices,  
$ \disp
 |{\widetilde p}  ^2  | = m^2  {  {\bf w}^2  \over  1  -  {\bf w} ^2  / c^2} $
where    ${\bf w}$ is the Newtonian velocity $ \disp {d {\bf x} \over dt }$.
Thus (\ref{sslow}) is satisfied 
provided $ \disp {\bf w}^2  / c^2   <  {1 \over  9}$, which 
corresponds to $|{\bf w}| < c / 3 $. 
Under this  limit, say  one third of the  velocity of light,
we  shall speak of a "moderately relativistic regime".

\medskip   \noi
  For inequal masses,  similar  results could be derived, but 
the discussion would become a bit complicated.      We summarize:

\noi  {\bf Proposition II}

\noi      {\sl In sofar as the mass differences are not too large, we
 keep the range of  $\sig, \epsilon$ under control by 
 restrictions  on the magnitude of the velocities. If in particular we consider 
three equal masses, velocities under $c/3$ ensure that we can invert 
our formulas with $S(\sig , \epsilon)$ as in Proposition~I.\/}

\noi   All the quantities involved in condition
 (\ref{slow}) (resp. (\ref{slowbis})) are first integrals for free 
particles, thus (\ref{slow})  (resp. (\ref{slowbis}))
defines an invariant sector of the motion.

\bigskip
\subsection{Individuality. New  {\em versus}  old coordinates}.

As a result of our transformation of the momenta, it might be puzzling that
 (beside its dependence on total momentum)   each  new variable  $q'_a$ 
depends not only on $q_a$ (with the same label $a$) but also on all $q_b$'s
 with $b\not= a$.                                            
This dependence is expressed  by the transformation formulas  (\ref{confi1}).
Fortunately,  we shall prove  that:

\noi {\bf Proposition III}

\noi   {\em   Beside its  dependence on the 
{\em direction} of $P$,  at zeroth order in $1/|P|$, the variable 
$z'_2$ depends only on $z_2$ (resp.  $ z'_3 $  depends only on $z_3$).   }

\noi    Proof

\noi    We develop our formulas in powers of $1/|P|$ and 
 evaluate $z'_{A \alp}$  at lowest  order. 

\noi According to  (\ref{confi1})
    we need to compute the coefficients 
$ \disp   {\partial y \over \partial  y'}$.

\noi               Let us first prove that
\beq   {\partial  y^\sig  _B    \over   \partial   y' _{A \alp}  }   =
 O (1/ |P|)  ,   \qquad \      { \rm for} \      A \not=B    
                                           \label{crossderiv}        \eeq
                   
\noi
 From (\ref{defxi}) and  (\ref{tild=})  
it is clear that 
\beq  y_2 ^ \alp   = 
  {\ytil}_2 ^{\prime \alp}  +  \   \xi \   P^ \alp ,  \qquad   \quad
      y_3 ^ \alp   = 
  {\ytil}_3  ^{\prime \alp}  +  \   \eta \   P^\alp  
                                        \label{-2-3}         \eeq
hence  
\beq   {\partial  y^\sig  _2    \over   \partial   y' _{3 \alp} }     = 
{\partial \xi      \over   \partial   y' _{3 \alp}  }   \   P^\sig  
                           \label {dy2/dy'3}               \eeq        
  \beq   {\partial  y^\sig  _2    \over   \partial   y' _{2 \alp} }     = 
\Pi  ^{\sig \alp}     +
{\partial \xi      \over   \partial   y' _{2 \alp}  }   \   P^\sig  
                           \label {dy2/dy'2}               \eeq         
 and similar formulas for $\dron y_3 / \dron y'_A $.
 We are  led  to evaluate the derivatives of $\xi$ (resp. $\eta$).
According to (\ref{etaxilin}) it is sufficient to differentiate $u$ and $v$.
With help of (\ref{ubis})(\ref{vbis}) we get
\beq P^2 {\partial  u    \over   \partial y'_{2 \alp} } =
P^\alp -   {\ytil}_3 ^\alp                        \label {1*}     \eeq
\beq P^2 {\partial  u    \over   \partial y'_{3 \alp} } =
\half  P^\alp -   {{\ytil}}_2 ^\alp  -    {{\ytil}}_3 ^\alp 
                                                      \label {un}     \eeq
\beq P^2 {\partial  v    \over   \partial y'_{2 \alp} } =
\half  P^\alp -   {\ytil}_2 ^\alp   -   {\ytil}_3 ^\alp  
                                                       \label {deux}   \eeq
\beq P^2 {\partial  v    \over   \partial y'_{3 \alp} } =
P^\alp -  {{\ytil}}_2 ^\alp                         \label {2*}     \eeq
Let us insert (\ref{un})(\ref{2*}) and (\ref{1*})(\ref{deux}) into the formulas 
obtained by  differentiation of (\ref{etaxilin}). We  obtain 
\beq   {\partial \xi      \over   \partial   y' _{3 \alp}  }    =
         O ({1/P^2})   \       \qquad  \qquad
 {\partial \eta   \over   \partial   y' _{2 \alp}  }    =
 O ({1/P^2})           \label{xi/3eta/2}         \eeq
\beq    {\partial \xi      \over   \partial   y' _{2 \alp}  }    =
{3\over 2}  {P^\alp  \over P^2}       +   O (1/P^2)        \label{xi/2}  \eeq
and a similar formula with  
$ \partial \eta    /   \partial   y' _{3}  $.         
Inserting       (\ref{xi/3eta/2})    into
 (\ref{dy2/dy'3})
  we check that   
$ \disp   {\partial  y^\sig  _2    \over   \partial   y' _{3 \alp} }   $ 
actually is   of the order of      $1/|P|$,  and  the same result can be 
derived  for   $\dron y_3  /  \dron y'_2$  ,   which altogether 
proves (\ref{crossderiv}).

\medskip
\noi  Now we apply formula  (\ref{confi1}) and take   (\ref{crossderiv})
      into account. Hence 
\beq {z'}_2    ^\alp    = 
{ \partial y_2 ^\sig   \over  \partial {y'} _{2  \alp } } \      z_{\sig 2}
+        0 (1/|P|)                                  \eeq
But in view of  (\ref{dy2/dy'2})(\ref{xi/2})  we simply have 
   $$ { \partial y^\sig  _2  \over    \partial y' _{2 \alp} }  = 
   \Pi ^{\sig \alp}    +   {3\over 2}  {P^\sig \   P^\alp \over  P^2}
+    O(1/|P|)        $$
So finally
\beq {z'}_2    ^\alp    = {\ztil}_ 2 ^\alp   +
   {3\over 2} \      { (z_2 \cdot  P)  \      P ^\alp   \over P^2}     +
 O (1/|P|)          \eeq                      
and a  similar  expression in terms of  $z'_3 ,  z_3$.
In particular we have
\beq {{\ztil}}'_2 = {\ztil}  _2  + O(1/|P|) ,  \qquad \quad
     {{\ztil}}'_3 = {\ztil} _3  + O(1/|P|)      \label{nomix}             \eeq

\bigskip
\subsection {New form of wave equation}

\noi   As seen in Section 4.1, the "difference equations" are (\ref{basic}) or 
equivalently (\ref{nu2})(\ref{nu3}).
\noi  According to (\ref{sumeq}) the dynamical equation (sum equation) 
for free particles is
\beq      ( 3 \sum m_a ^2  c^2    -P^2 ) \   \Psi  =
(D + 6 P^2 \Xi )  \   \Psi                     \label{newfree}     \eeq
Of course $\Xi$ must be here considered as a function of 
$   y' _2 \cdot P  , \  y'_3 \cdot P, \   P^2 ,  \  \ytil  _A  $.
In view of (\ref{Xinew})(\ref{igrec})  we can write  as well
\begin{equation} \Xi = {3 \over 4} X^2  +  {\epsilon ^2  \over (2- 3X) ^2}    
  \label{XideX} \end{equation}
where $X = S (\sig , \epsilon)$ according to (\ref{dvlop})(\ref{dvlopS}).
We must remember that $\sig , \epsilon$ are  functions of the new 
momenta      through (\ref{uv}).

\noi  But equation        (\ref{basic})   tells that  
{\em on the mass shell} we can  replace  $  y' _{1A}   \cdot P $
by $  \nu _A  c^2 $ (thus $y'_A$  replaced accordingly, see eqs. 
(\ref{nu2}) (\ref{nu3})  ).
Moreover we impose that the total linear momentum has a sharp value $k^\alp$.
Let us make this convention that $\underline F$ is the expression of any $F$ 
on the momentum-mass shell, namely
\beq  {\underline F} = {\rm subs} \    (y'_{1A} \cdot P = \nu _A c^2 ,
   \qquad      P^\alp =  k^\alp , \qquad      F)   \eeq
using an  obvious  notation borrowed from Maple's syntaxis.  It is meant that
$y'_{1A} $  is  as in (\ref{difp'})   and    we set
\beq        k^2  =  M^2  c^2              \label{defM}     \eeq
For instance,  if we define
$$  {{\yhat}'}_\alp  =  {y' }_\alp  -  (y' \cdot k / k^2 )  k_\alp        $$
we can write
  $ {\underline  \ytil }^\alp    =   { \yhat} ^ \alp  $,  therefore
$${ \soulD} =  6 [(\yhat ' _2)^2  +  (\yhat ' _3 )^2   +
\yhat '_2   \cdot \yhat ' _3    ]                                  $$ 
Moreover   (\ref{uv})  yields

\beq  M^2  c^2  \soulu = \nu_2 c^2 
      - ( \yhat_2 \cdot \yhat_3 + \half \yhat_3 ^2)
     \qquad \quad
 M^2 c^2  \soulv  = \nu_3  c^2 
                - ( \yhat_2 \cdot \yhat_3 + \half \yhat_2 ^2)    
                                                \label{soulu,v}       \eeq
It is noteworthy that, in the case of {\sl two equal masses}
 ${\underline  \epsilon} $ 
is of the    order        of    $ \disp    1 / c^2$,  whereas for {\sl three 
equal masses} both         
$ {\underline \sig}$ and  ${\underline \epsilon}$
  are  $ \disp  O (1 / c^2) $.

\noi  Taking into account the mass-shell constraints and the sharp value of 
$P^\alp$ we derive  the reduced equation  
\begin{equation}  ( 3 \sum m_a ^2  -M^2 ) c^2   \   \psi  =
(\soulD    +      6  M^2  c^2    \  {\underline \Xi})   \     \psi
                                  \label{final}       \end{equation}
Notice that, apart from $ \nu_2  ,  \nu_3$ that are fixed  parameters, 
 $\underline \Xi$  depends only on  
    $   \yhat ' _2  , \quad     \yhat ' _3 $  and  $M^2$.
The only operators involved in (\ref{final}) 
are multiplications by  the projections of  $ y'_A $ 
 orthogonal to  $k$, they are essentially three-dimensional.
Whereas $\soulD$ has a familiar form (just use the rest frame,  where
 $\yhat_A  \cdot  \yhat _B  = - {\bf y}_A  \cdot  {\bf y}_B $) it is not the 
case for ${\underline \Xi}$.
Fortunately it can be checked that, at least for equal masses,  the term 
    $c^2 {\underline \Xi}$ is in fact of the order of $1/c^2$.
For this purpose it is convenient to set
\beq    M^2 c^2    {\underline \Xi}  = 
 {1  \over  M^2 c^2 }  \       \Gam       \label{defGam}     \eeq
so we end up  with 
\begin{equation}  ( 3  \sum  m_a  ^2    -M^2 ) c^2  \   \psi  =
{\underline D}    \      \psi
+ {6 \over M^2  c^2}  \Gam    \      \psi   \label{finalred}     \end{equation}

 {\sl For three equal masses}, $\Gam$ can be expanded in non-negative
powers of $1/ c^2$ and it turns out that its  zeroth-order piece is biquadratic 
in ${\yhat}' _A$.

{\sl Proof}. It can be easily read off from  (\ref{uv}) that in this case 
$ \soulu ,  \soulv $  thus also  $ \soulsig ,  \souleps $
 are of the order of   $1/c^2$.
Getting back to  system (\ref{linxi})(\ref{lineta})   one finds that
$$ {\underline \xi} = 2 \soulu - \soulv   +  O (1/c^4)  $$ 
$$ {\underline \eta} = 2 \soulv - \soulu   +  O (1/c^4)  $$
Inserting into  (\ref{defXi})  yields
\beq   {\underline \Xi} = 3 (\soulu ^2  +  \soulv ^2 - \soulu  \soulv )
                +  O (1/c^4)                                 \eeq
hence
\beq  M^4 c^4  {\underline \Xi} =   \Gam (0) +  O (1/c^2)         \eeq
\beq   \Gam (0)  =
  {3 \over 4}   \        [
  (\yhat _2 ^2 )^2   +   (\yhat _3 ^2 )^2
+     4   ( \yhat _2  \cdot  \yhat _ 3 )^2
+  2  (  \yhat _2 ^2   +  \yhat _3 ^2  )  \    (\yhat _2  \cdot \yhat _ 3)
-  (\yhat _2 ^2 )   (\yhat _3 ^2 )           ]
                                      \label {biqadr}     \eeq
Thus, when all $m_a = m$,
 the last term in the r.h.s. of  (\ref{finalred})  can be considered as small.
        
\medskip
\noi  Free-particle motion is  now  described  only  in terms of 
$ {\widehat y} '$ and $k$.
  
\medskip
\noi                     Imposing by (\ref{P}) that total linear
 momentum is diagonal permits, through   equation (\ref{psi}),
 to eliminate    $y'_A \cdot k$, 
where  the new relative energies   $c  y'_A \cdot k / \sqrt {k^2}$
are  conjugate to the  new "relative times". 

It is of interest to notice that these  new "relative times"
are linear combinations of  the old ones with coefficients that are analytic 
functions of the momenta; the reader will check it using 
(\ref{confi1})(\ref{confi2})(\ref{defxi}) 
and (\ref{uv})(\ref{rhs2})(\ref{rhs3}).  

  After reduction, the three-body kinematics has no        
 more degrees of freedom than in the non-relativistic problem.
But we must keep in mind that this 
picture is valid only in sofar as we can revert to all the  initial variables,
which (at least for equal masses) is ensured for moderately relativistic 
 velocities.

\noi   The new variables $y'_A$ introduced in this Section   will be referred 
to as the {\em reducible  variables\/}.

\bigskip
\section {How to introduce interactions}

\noi We can  now consider the system  (\ref{newfree})
 (\ref{nu2})(\ref{nu3})
 as a starting point for  introducing mutual interactions.

\noi To this end, we shall modify the "sum equation" (\ref{newfree})
 by a term which carries  interaction,
 whereas  the "difference equations" (\ref{nu2})(\ref{nu3}) 
remain untouched.

\noi
 Doing so  we manage that $P$ remains conserved, and keep assuming that
its  eigenvalue  is a timelike vector $k$;
 therefore the factorization of $\Psi$ given by 
formula (\ref{psi}) remains valid and eliminates two degrees of freedom.

\bigskip
\noi 
The interaction potential  will be  written in closed form in terms of
 the reducible coordinates   $z' _A , y' _B $, and all calculations will be 
carried out using these variables. 

\noi Remark:  the {\em reducible (momentum)  coordinates\/} 
 $p_a '$ are re-arranged   as  to form the quantities
 $P$ and $y'_A$.

\bigskip
\noi  Adding interaction into  (\ref{newfree}) produces  the  dynamical equation
\begin{equation} ( 3 \sum m_a ^2  c^2      - P^2) \Psi=   
 D  \  \Psi
+  (18 V + 6 P^2  \Xi ) \  \Psi             \label{wexum}     \end{equation}
Like in the free case,
 $D$ is given by (\ref{defD}) and $\Xi$ is given by (\ref{XideX}) 
in terms of $X=  S(\sig, \epsilon)$.

\noi     The "difference equations" 
 remain (\ref{nu2})(\ref{nu3}) like previously.
 Of course, $V$ cannot be chosen 
arbitrarily but it is not difficult to find a general admissible form of $V$
such that the dynamical equation (\ref{wexum})
 is compatible with (\ref{nu2})(\ref{nu3}).       Compatibility requires 
that $V$ commutes with the operators in the left-hand sides of
 (\ref{nu2})(\ref{nu3}).      
For instance  the interaction potential $V$ may depend  on 
${\widetilde z}'_2 , \         {\widetilde z}' _3  $  and $P^2$.

\noi Naturally $V$ must be Poincar\'e invariant, which is realized by taking a 
function of the various scalar products formed with 
${\widetilde z}'_A , \ytil '_B , P$.

\noi             Demanding that $\Psi$ diagonalizes 
$P^\alpha$  with eigenvalue $k^\alpha$, with $k\cdot k >0$, we can in 
 (\ref{wexum})    replace  $\ytil '$ by ${\widehat y}'$.

\noi   Taking  (\ref{nu2})(\ref{nu3}) and  (\ref{defGam})
into account yields the reduced equation
\begin{equation}  ( 3  \sum m_a ^2   -  M^2) c^2     \   \psi  =
\soulD \psi    + 18  \soulV   \   \psi
+ { 6 \over  M^2 c^2 }    \Gam      \  \psi   
                                              \label{pdv}  \end{equation}
where  the reduced wave function
$\psi$ depends only on $k$ and on the space projections  
${\yhat} '_2, \  {\yhat} '_3 \  \  $.
The only operators involved here  are the projections 
 $\zhat ' _A , \   \yhat '_B $.    
 Moreover  $ {\zhat}' $ arises in  $\underline V$ only.

\noi   Comparison with a standard problem of nonrelativistic quantum mechanics
becomes more easy in the rest frame, where 
$({\zhat}' _A ) ^2 = - ({\bf z}'_A )  ^2     $  and 
$({\yhat} ' _A ) ^2 = - ({\bf y}' _A )   ^2                $, etc.

\noi
Actually   solving   (\ref{pdv}) differs from  a 
 non-relativistic  problem by  the last  term,  which involves the 
momenta  but does not depend on the shape of the interaction (and survives in 
the free-motion limit).  Still this term depends on  the total squared mass.

\noi      For simplicity, we can consider  an  interaction  such that
\beq  18 V =      \alp _{12}  \   U _ {12}  ({\ztil} '_2 )       +
             \alp _{23}   \   U _ {23} ({\ztil} '_3  -  {\ztil}'_2)
+      \alp _{31}    \    U _ {31}   ({\ztil}'_3)    
                                                          \label{18V}   \eeq
where $\alp _{ab} $ are  coupling constants  and  $U_{ab}$ arbitrary 
(but Poincar\'e invariant) functions.
In this model, $U_{12} $ is independent from $q'_3$, etc,  with cyclic 
permutation; the formal input of our interaction consists in two-body  
potentials.

\noi    So (\ref{pdv})  can be written 
$$  (3  \sum m_a ^2 - M^2) c^2   \psi = \qquad  \qquad   \qquad              $$
\beq        \soulD  \psi     + 
 [ \alp _{12}  \     U_{12}  ( {\zhat }' _2 ) +
\alp _{23}   \     U_{23}  ( {\zhat }' _3  -    {\zhat }' _2  )      +
\alp _{31}   \     U_{31}  ( {\zhat }' _3 )   ] \     \psi
+  {6 \over  M^2 c^2}  \Gam   \    \psi     \label{pdv2}   \eeq
A special case 
\beq  U_{12}  =    ( {\ztil} '_2 )^2    , \qquad
      U_{13} =     ({\ztil} '_3  ) ^2   , \qquad
U_{31}   =  ( {\ztil} '_3  -  {\ztil} '_2 )^2        \label{osc}        \eeq    
 describes a three-boson harmonic oscillator.

\noi In order to handle equation (\ref{pdv}) it is tempting to neglect its 
 last term. Invoking the limit of a large total momentum ($M^2 \rightarrow 
\infty$ ),  as in \cite{houch}, doesnot seem to  permit a 
perturbation treatment. We prefer to consider developments in powers of 
$1/c$.

\medskip
\subsection{Equal masses}

\noi  Assuming for simplicity that $m_a = m$, equation (\ref{pdv}) becomes 
\beq  (9m^2 -M^2) c^2  \   \psi =
\soulD \psi    + 18  \soulV   \   \psi
+ { 6 \over  M^2 c^2 }    \Gam      \  \psi   \label{pdvm}       \end{equation}
It will be considered as an eigenvalue problem for $\lam$
by   setting
   $ 6 \lam =  (M^2 - 9m^2) c^2  $.  As all masses are equal,   thus
 $\Gam = \Gam (0) + 0 (1/c^2)$. At first order in $1/c^2$ we can replace 
$\Gam$ by $\Gam (0)$ and  $M^2$ by $9m^2$ in           (\ref{pdvm}). Using the 
rest frame we obtain
\beq   \lam  \psi=
 [ { \bf  y} _2  ^2   +   { \bf y} _3  ^2   +  {\bf y}_2  \cdot  {\bf y}_3  
       -3  \soulV   ] \     \psi
   - {1 \over  9m^2   c^2 }   \Gam (0)  \    \psi    \label{approx}    \eeq
Neglecting the last term yields the nonrelativistic limit (divide by $m$ and 
remember that in our formulas, $V$ has dimension of $P^2$).

\noi Taking into acount the    contribution of $\Gam (0)$ permits to calculate 
the first relativistic correction.

\medskip
\subsection{Cluster behaviour}

\noi
As pointed out by Sazdjian \cite{Saz1}, in any formulation of the dynamics
 which makes explicit reference to total momentum, it is  difficult to
 discuss cluster separability.  But it is reasonable to demand 
 that the {\em reduced} equation be in a sense separable, in order to 
ensure a factorization of the internal wave function when there are 
noninteracting clusters. 

\noi     With this requirement in mind,  we can already   observe that the
 potential  (\ref{18V}) is    formally separable
 {\em in terms of the variables}   $z'$.  

\noi
But the interpretation of each $U_{ab}$ as a two-body term  runs into 
 a complication:  there is no evidence   that the  variable
 $z'_A$,  exactly matches   the  cluster of   particles \{1A\}. 
A similar remark arises  concerning   the matching of $z'_3 - z' _2$
 with  cluster \{23\}.  

\noi
The physical interpretation of the new configuration variables $z'_A$ is not 
straightforward;  they  are relative variables since they commute with $P$,
but they  suffer from this complication that the
 transformation formulas  (\ref{confi1})   mix $z_2$ with $z_3$.
 Similarly  (beside its dependence on total momentum)   each  new variable
  $q'_a$ depends not only on $q_a$ (with the same label $a$)
 but also on all $q_b$'s   with $b\not= a$.

But we can consider (\ref{nomix}) on the momentum-mass shell. At least for 
three equal masses, the only occurence of the velocity of light is through  
the product  $Mc$, so we obtain from (\ref{nomix})
\beq {{\zhat}}'_2 = {\zhat}  _2  + O(1/|Mc|) ,  \qquad \quad
     {{\zhat}}'_3 = {\zhat} _3  + O(1/|Mc|)                   \eeq
and, of course, 
$  {{\zhat}}'_3 - {{\zhat}} ' _2  =  \zhat _3  -  \zhat _2    + O(1/|Mc|) $.
Thus, at leading  order, the variables ${{\zhat}}'_A$ and $\zhat _A$
still  coincide; so
the potentials  $U_{ab} $  in the reduced equation
     (\ref{pdv2})  can be   approximately considered as  two-body terms.

            $$                  $$                                       

\section  {Concluding remarks}

\noi
As a first step, we succeeded in  constructing  three mass-shell constraints 
describing  the {\sl free motion}  of three scalar particles.
In contrast to the KG system, these new wave equations permit to eliminate two 
degrees of freedom and get reduced to a covariant equation with 
three-dimensional arguments.

\noi Our approach rests on a transformation of the momenta involved 
in the original KG system.
In contradistinction to Sazdjian's proposal  and the 
 homographic relations that approximate it 
(eq (13) of ref.  \cite{Saz1},    eq (4.15) of ref. \cite{Saz2}),
our transformation from the old momenta to the new ones is explicitly given
 by simple {\em quadratic\/} formulas.                       
Inversion of these formulas is a fourth degree algebraic problem which could be 
(in principle) discussed and solved in closed form;  due to its complexity, 
 approximate developments are  more efficient  in practical calculations.

\noi  We used a couple of identities that  are specific of the
 three-body case; thus  an
extension of the present work to $n> 3$ is by no means straightforward!

\noi    In the present state of the art, equivalence of the new equation
(\ref{final})  with the sum of the original KG equations 
 is ensured {\em at least} in a large sector  characterized by 
 positive energies and conditions that involve the masses of the particles.
When the masses are not too different one from another (and in particular for 
equal masses), these conditions  amount  to impose a bound on the velocities;
but this bound is still high enough to allow  for the description of 
 a  relativistic regime.

\noi   The case of very large velocities requires further investigations.
 We gave 
here {\em sufficient} conditions for an invertible transformation;
 it remains possible that a more detailed discussion enlarges the present 
results.

\noi This analysis of free-body kinematics provides us with a solid ground.

\medskip
\noi
In a second step, we  introduced  {\sl interaction\/} in the "sum equation".
 The model obtained by this procedure  respects Poincar\'e invariance.
It remains  covariantly reducible to a wave equation with three-dimensional 
arguments; 
free motion is recovered in the absence of interaction term.

 The interaction term  $V$  is {\em formally} cluster separable; 
actually formula   (\ref{18V})  is an ansatz which permits to combine two-body 
interactions without spoiling the  compatibility of the mass-shell 
constraints. True separability  (in terms of the original individual
 particle coordinates) is recovered  only in the large-total-mass limit.

\noi  The two-body input of our model can be either phenomenological or 
motivated by consideration of  field theory.

When  the three masses are equal, the velocity of light 
 arises in  $\underline \Xi$ through the product 
 $M^2 c^2$, which facilitates the expansion in powers of  $1/c^2$.
At the first order,  the reduced equation is similar to  a familiar 
Schroedinger equation supplemented with  a perturbation;
 insofar as the interaction is not explicitly energy dependent (or if 
this dependence is of higher order)
one is left with  a conventional eigenvalue problem.

\noi      This situation provides a basis for eventually  undertaking
 the study of cases  where the mass differences are not zero but
 still remain relatively small.

\noi
In the hope of  applications to three-quark or three-nucleons systems,
 we plan an extension of the formalism to particles with spin. The contact with 
  more elaborated (but more complicated) theories, such as  QED and QCD,
  will be discussed in a future work.

\bigskip

\noi  {\bf Appendix 1} 

\medskip

\noi   In non-relativistic classical mechanics, canonical transformations 
are symplectic diffeomorphisms of phase space.
In general they  do mix  the $q$'s and the $p$'s.

\noi But a {\em point transformation\/} (in configuration space) simply  
transforms   the  $q$'s among themselves, say  $q'=  f(q)$.
 Then invariance of the symplectic form \cite{gold}
fully determines the $p'$'s in terms of {\em the $q$'s and the \/} $p$'s.
When configuration space is flat, the $q$ and $p$ variables play symmetric 
roles in the general formulas of analytic mechanics, so there is no 
difficulty in defining  as well 
 point transformations in momentum  space (but this possibility is 
not usually  considered in textbooks).  
      In this case, one  transforms  the momenta among 
themselves, and one further determines  the new  variables  $q'$
 in terms of $q$'s and $p$'s through the 
requirement that the complete transformation law is canonical. 
 
\noi   In the position (resp. momentum) representation of quantum mechanics, a
quantum analog of point transformations in configuration (resp. momentum) 
space can be generated by an invertible transformation of the arguments of 
the wave function.
This transformation among $c$-numbers obviously induces a transformation among 
the multiplicative operators they define.

\medskip
\noi {\bf  Appendix 2.}

 The polynomial $R_\sig (X)$ has an obvious double root $X = 2/3 $ 
independent of $\sig$, and 
 provided $\sig > -1$, 
 two other real roots given by (\ref{Xpm}) but, as noticed 
above, the root $X^-$ falls outside the admissible interval. 

\noi    All the curves $Z= R_\sig (X)$ are tangent to the $X$ axis at a
 fixed point
$X= 2/3$.                For  $\sig > -1$  and $\epsilon$ small enough,
 the curve representing $R(X)$ is four times cut by  the straight line 
$Z= \epsilon ^2$.  
In the limit when $\epsilon$ vanishes,
 two points of this intersection form the contact with the $X$ axis, and 
the  other ones respectively reduce to $X^+$ and $X^-$. 

\noi   For $\sig = \half$ we find that $X^+   \simeq  0.15$, which is 
admissible in the sense of   (\ref{ineq1})(\ref{ineq2}),  and  
$R_{\half} (X)$   has a local maximum  at   $X= 1/3$. This maximum is exactly 
${1\over 4}$.
For $\sig < \half$, the local maximum exceeds ${1\over 4}$.  Making $\sig$ to 
decrease we obtain lower values of $X^+$ (which  vanishes with $\sig$). 

\noi  For   $\sig = -{3\over 4}$,  we obtain exactly   $X^+ =  -{1\over 3} $, 
and going down further is excluded in view of (\ref{ineq2}).

\noi Taking $\sig$ in the open interval $ ( -{3\over 4} ,  \half )$
and $X$  restricted by $  -{1\over 3} < X  < {2 \over 3} $, 
it turns out that, {\em provided} $\epsilon ^2 < {1 \over 4} $,
each  curve $Z = R_\sig  (X) $  has two points in common 
with the straight line  $Z = \epsilon ^2$ (other possible points correspond 
to $X$ outside the interval we consider).
For vanishing $\epsilon$, 
one of them has its horizontal coordinate  going to coincide  with $X^ +$,  
while  the other point goes to the fixed contact point $X= 2/3, Z= 0$.
This analysis shows  that, with our restrictions,   the 4th dgree equation 
  $R_\sig (X) = \epsilon^2$ has two  real solution, but only one of them 
reduces to $X^+$ in the limit where $\epsilon$ vanishes.


\begin{thebibliography}{3}
\bibitem{JGTaylor}  J.G. Taylor, Phys. Rev. {\bf 150}, 1321 (1966).\\
 A. Pagnamenta,
 in "Lectures in Theoretical  Physics" Univ. of Colorado, 
Boulder, Mathematical Methods in Theoretical Physics,
K.T. Mahanthappa and W.E. Brittin Eds, Gordon and Breach, New York (1969).\\
T.L. Basdevant and R. Omn\`es,
 Phys. Rev. Letters {\bf 17}, 775 (1966).

\bibitem{Tjo} G. Rupp, J.A. Tjon,   Phys. Rev. {\bf C 45},2133 (1992).\\
F. Sammarruca, D.P. Xu,   R. Machleidt,   Phys. Rev. {\bf C 46}, 1636 
(1992).\\
A. Stadler, F. Gross , Phys. Rev. Lett.{\bf 78},26 (1997).

\bibitem{Bij} J. Bijtebier,
Two and three-fermion 3D equations deduced from  Bethe-Salpeter equations,
hep-th/9912099.  

 J. Phys. G {\bf 26} (2000), 871-886. 
              

\bibitem{predic1}
"Relativistic Action-at-a-distance Classical and quantum aspects", Lecture
Notes in Phys. {\bf 162}, J.Llosa Editor, Springer Verlag (1982) and references
therein.   


\bibitem{predic2}  Ph. Droz-Vincent,  Reports in Math. Phys. {\bf 8}, (1975) 79;
 Phys.Rev.D {\bf 19}, 702 (1979).\\    
I.T. Todorov, JINR Report E2-10125, unpublished (1976).\\
L.Bel,  in Differential Geometry and Relativity, M.Cahen and M.Flato editors,
Reidel Dordrecht (1976) 197, Phys. Rev. D {\bf 28} (1983) 1308.\\
  H. Leutwyler and J. Stern, Ann.of Phys.(N-Y){\bf 112} (1978) 94, Phys.Lett.B
 {\bf 73} (1978) 75. 

\bibitem{predic3}   
V.A. Rizov, H.Sazdjian and I.T. Todorov,  Ann.of Phys. {\bf 165}, 59 (1985).
H.Sazdjian,   Phys.Lett.B {\bf 156}, 381 (1985);   Jour.Math.Phys. {\bf 28},
 2618 (1987).

\bibitem{llos}    V. Iranzo, J. Llosa, F. Marqu\'es and A. Molina,
Ann. Inst. H. Poincar\'e, {\bf A 35}, 1 (1981);            
 {\bf A 40}, 1 (1984).
 
 \bibitem{DV} Ph. Droz-Vincent Phys. Letters {\bf 159 B}, 
393 (1985).   In spite of the presence of $D$, the 3-body wave equation
 proposed  therein   cannot be identified with eq. (\ref{pdv})
  of the present work. This remark also applies to  \cite{llos}.
             

\bibitem{bijNCim}   J. Bijtebier, 
Three-fermion relativistic bound states and cluster decomposition
Nuovo Cim. {\bf A 103} 669-682 (1990).

\bibitem{Saz1}     H. Sazdjian,      
 N-Body bound state relativistic wave equations
Annals of Phys.       {\bf 191}, 52-77 (1989). 

\bibitem{Saz2}     H. Sazdjian,
 Relativistic dynamics for N-body systems,
 Physics Lett. {\bf B 208}, 470-474 (1988); 
   

\bibitem{houch}  Ph. Droz-Vincent, Few-Body Systems {\bf 31}, 165-170 (2002).
Several coefficients in the expression of $\Gam (0)$ written in that 
article must be corrected, see eq.  (\ref{biqadr})  in the present paper.   

\bibitem{gold}   H. Goldstein, "Classical Mechanics" Chap. 8, section 8-2.
 Addison-Wesley, Reading (Mass.) (1964)

\end{thebibliography}
\end{document}